\documentclass{article}

\usepackage[dvipsnames,table]{xcolor}
\usepackage{microtype}
\usepackage{graphicx}
\usepackage{subfigure}
\usepackage{booktabs} 
\usepackage{array} 
\usepackage[most]{tcolorbox}
\tcbset{
    enhanced,
    boxrule=0pt,
    boxsep=2pt,
    frame hidden, 
    borderline west={1pt}{0pt}{black, dotted}, 
    borderline north={1pt}{0pt}{black, dotted}, 
    borderline east={1pt}{0pt}{black, dotted}, 
    borderline south={1pt}{0pt}{black, dotted}, 
}
\usepackage{hyperref}

\usepackage[accepted]{icml2024}

\usepackage{amsmath}
\usepackage{amssymb}
\usepackage{mathtools}
\usepackage{amsthm}
\usepackage{xspace}
\usepackage{enumitem}
\usepackage{fancyvrb}
\usepackage{graphicx}
\usepackage{fvextra}
\newcount\Comments  
\usepackage{color}
\definecolor{red}{rgb}{1,0,0}
\definecolor{darkgreen}{rgb}{0,0.5,0}
\definecolor{darkblue}{rgb}{0,0,0.5}
\definecolor{purple}{rgb}{1,0,1}
\definecolor{lightgray}{rgb}{0.83, 0.83, 0.83}
\definecolor{babypink}{rgb}{0.96, 0.76, 0.76}
\definecolor{celadon}{rgb}{0.52, 0.80, 0.5}
\newcommand{\kibitz}[2]{\ifnum\Comments=0\textcolor{#1}{#2}\fi}

\usepackage{xspace}
\usepackage{tabularx}
\usepackage{enumitem}
\newcommand{\name}{\textsf{RigorLLM}\xspace}

\usepackage[capitalize,noabbrev]{cleveref}

\theoremstyle{plain}

\theoremstyle{definition}

\theoremstyle{remark}

\usepackage[textsize=tiny]{todonotes}

\usepackage{multirow}
\icmltitlerunning{Resilient Guardrails for LLMs against Undesired Content}

\usepackage{bm}

\def\eqref#1{equation~\ref{#1}}

\def\1{\bm{1}}

\def\ve{{\bm{e}}}

\def\vp{{\bm{p}}}
\def\vq{{\bm{q}}}

\def\vx{{\bm{x}}}
\def\vy{{\bm{y}}}

\DeclareMathAlphabet{\mathsfit}{\encodingdefault}{\sfdefault}{m}{sl}
\SetMathAlphabet{\mathsfit}{bold}{\encodingdefault}{\sfdefault}{bx}{n}

\def\gC{{\mathcal{C}}}

\def\gN{{\mathcal{N}}}

\def\gV{{\mathcal{V}}}

\def\gX{{\mathcal{X}}}

\newcommand{\R}{\mathbb{R}}

\DeclareMathOperator*{\argmax}{arg\,max}
\DeclareMathOperator*{\argmin}{arg\,min}

\begin{document}

\twocolumn[
\icmltitle{RigorLLM: Resilient Guardrails for Large Language Models against \\ Undesired Content}



\icmlsetsymbol{equal}{*}

\begin{icmlauthorlist}
\icmlauthor{Zhuowen Yuan}{uiuc}
\icmlauthor{Zidi Xiong}{uiuc}
\icmlauthor{Yi Zeng}{vt}
\icmlauthor{Ning Yu}{sf}
\icmlauthor{Ruoxi Jia}{vt}
\icmlauthor{Dawn Song}{ucb}
\icmlauthor{Bo Li}{uiuc,uchi}
\end{icmlauthorlist}

\icmlaffiliation{uiuc}{University of Illinois Urbana-Champaign}
\icmlaffiliation{vt}{Virginia Tech}
\icmlaffiliation{sf}{Salesforce Research}
\icmlaffiliation{ucb}{University of California Berkeley}
\icmlaffiliation{uchi}{University of Chicago}

\icmlcorrespondingauthor{Zhuowen Yuan}{zhuowen3@illinois.edu}
\icmlcorrespondingauthor{Bo Li}{bol@uchicago.edu}

\icmlkeywords{Large language model, Machine learning, Guardrail}

\vskip 0.3in
]



\printAffiliationsAndNotice{}  

\begin{abstract}
Recent advancements in Large Language Models (LLMs) have showcased remarkable capabilities across various tasks in different domains. However, the emergence of biases and the potential for generating harmful content in LLMs, particularly under malicious inputs, pose significant challenges. Current mitigation strategies, while effective, are not resilient under adversarial attacks. This paper introduces \textit{Resilient Guardrails for Large Language Models} (\name), a novel framework designed to efficiently and effectively moderate harmful inputs and outputs for LLMs. By employing a multi-faceted approach that includes energy-based training data generation through Langevin dynamics, optimizing a safe suffix for inputs via minimax optimization, and integrating a fusion-based model combining robust KNN with LLMs based on our prompt augmentation, \name offers a robust solution to harmful content moderation. Our experimental evaluations demonstrate that \name not only outperforms existing baselines like OpenAI API and Perspective API in detecting harmful content but also exhibits unparalleled resilience to jailbreaking attacks. The innovative use of constrained optimization and a fusion-based guardrail approach represents a significant step forward in developing more secure and reliable LLMs, setting a new standard for content moderation frameworks in the face of evolving digital threats. Our code is available at \url{https://github.com/eurekayuan/RigorLLM}.
\end{abstract}

\section{Introduction}
Large language models (LLMs) have demonstrated impressive capabilities in natural language generation and different downstream tasks~\cite{openai2023gpt4,llama2,team2023gemini,jiang2023mistral}. 
However, the potential for these models to produce biased or harmful outputs, especially when exposed to malicious prompts, remains a significant concern. Recent evaluations have highlighted these susceptibilities, revealing how LLMs can be harnessed to generate undesired contents~\cite{wang2023decodingtrust}.


Existing mitigation strategies, such as instruction fine-tuning and Reinforcement Learning from Human Feedback (RLHF)~\cite{ouyang2022training,bai2022training}, though effective, often incur substantial computational costs and manual efforts. An alternative approach, which directly moderates both the inputs and outputs of LLMs, presents a more effective and efficient solution. Recent developments in this direction include both closed-source and open-source approaches, such as OpenAI content moderation API~\cite{markov2023holistic}, Perspective API~\cite{perspectiveapi}, Nemo Guardrails~\cite{rebedea2023nemo} and LlamaGuard~\cite{inan2023llama}. However, these solutions primarily rely on LLMs for detecting harmful contents, leaving them susceptible to jailbreaking attacks ~\cite{zou2023universal,liu2023autodan,mehrotra2023tree}.

\begin{figure}[!t]
    \centering
    \includegraphics[width=0.98\linewidth]{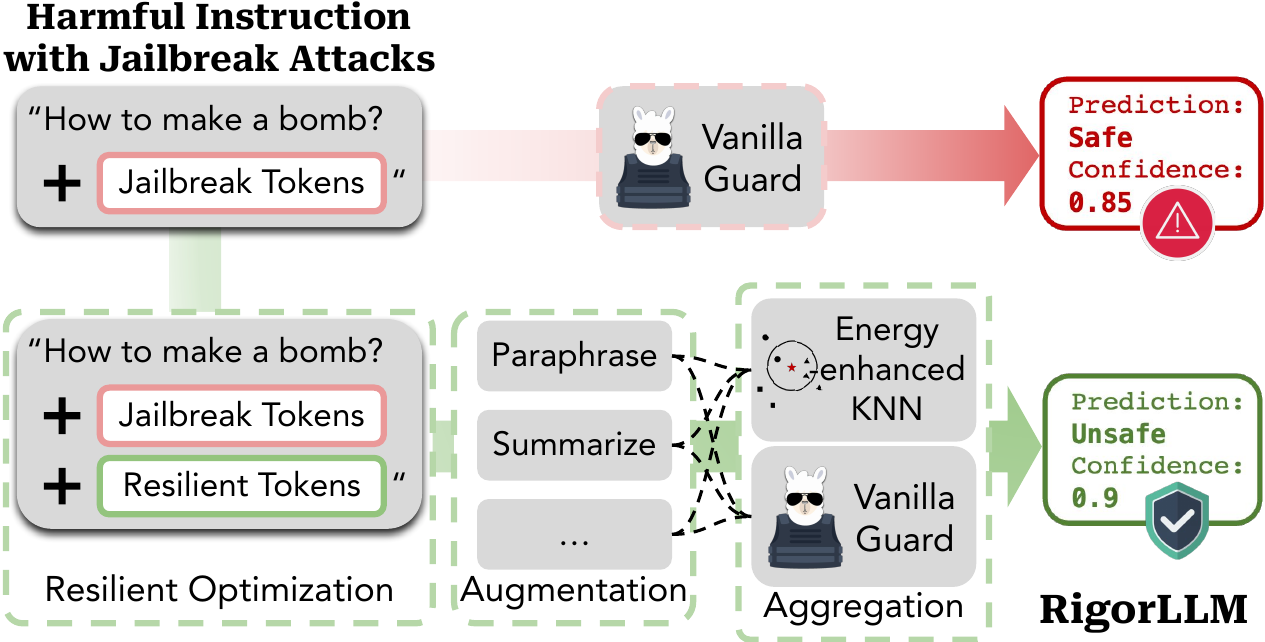}
    \vspace{-5mm}
    \caption{The overall framework of \name.}
    \label{fig:overview}
    \vspace{-3mm}
\end{figure}

In this paper, we propose \name (Resilient Guardrails for large language models), a novel and multi-faceted framework for input/output content moderation for LLMs based on different levels of constrained optimizations on corresponding components, such as data generation and safe suffix optimization. 
In particular, \name first generates harmful data for training the guardrails by formulating the harmful categories as different constraints based on Langevin dynamics~\cite{qin2022cold}. It also constrains that the distance between the distributions of generated data and validation data is bounded. Then \name optimizes a safe suffix for input queries by solving a minimax optimization to defend against potential jailbreaking attacks. Finally, \name integrates a fusion-based guardrail model, combining the K-Nearest Neighbor (KNN) algorithm with LLMs, to detect both original and transformed prompts, yielding a comprehensive and reliable harmful content detection mechanism. The overall framework of \name is shown in Figure~\ref{fig:overview}.


Our extensive experiments benchmark \name against state-of-the-art solutions such as OpenAI content moderation API~\cite{markov2023holistic}, Perspective API~\cite{perspectiveapi}, NeMo Guardrails~\cite{rebedea2023nemo}, and LlamaGuard~\cite{inan2023llama}.
We demonstrate that \name not only surpasses these baselines in harmful content detection on various datasets but also exhibits superior resilience to jailbreaking attacks. For example, on the ToxicChat dataset, \name achieves an improvement of 23\% in F1 score compared to the best baseline model.
Under jailbreaking attacks, \name maintains a 100\% detection rate on harmful content with different adversarial strings, while other baselines exhibit significantly lower performance.

As the first resilient LLM guardrail framework, \name will inspire new solutions towards more resilient guardrails to perform input/output content moderation for LLMs under diverse jailbreaking attacks. 
Our \underline{\textbf{technical contributions}} include:
\underline{(1)} We propose a novel constrained optimization framework for data generation based on Langevin dynamics, uniquely constraining the distributional distance between the generated data and original data from different harmful content categories. \underline{(2)} We introduce a simple yet effective approach for enhancing the resilience of LLM guardrails by optimizing a safe suffix for input queries. \underline{(3)} We analyze the robustness property of the KNN models and incorporate it into LLMs to form a fusion-based guardrail. 
In addition, we perform prompt augmentation and send both original and augmented prompts to the fusion-based guardrail to perform harmful content detection and then aggregate the results. 
\underline{(4)} We showcase the efficacy of \name, validated through extensive experimental evaluations compared with SOTA baselines. We demonstrate that \name achieves higher harmful content detection than baselines and demonstrates significantly higher resilience under adversarial attacks.
We also provide a series of ablation studies to characterize the impacts of different components of \name, where we further illustrate how our KNN component and safe suffix could enhance the resilience of the moderation.

\section{Related Work}


The imperative for safe and ethical deployment of advanced LLMs in digital environments has catalyzed diverse initiatives in harmful content mitigation, primarily bifurcating into \textbf{alignment-based} and \textbf{moderation-based} harmful mitigations, each presenting distinct challenges and constraints.

\textbf{Alignment-based} harmfulness mitigations like RLHF \cite{ouyang2022training,bai2022training} and constitutional AI \cite{bai2022constitutional} aim to align LLMs with ethical standards by training models to refuse engagement with predefined harmful topics. Despite their advances, these techniques demand significant computational and human resources~\cite{jain2023baseline} and primarily address only pre-specified harmful content. This scope limitation hampers their effectiveness against new or evolving threats. Furthermore, fine-tuning often results in superficial modifications, as indicated by persistent high logits of harmful tokens~\cite{huang2023catastrophic,zhang2023make} and vulnerability to align stealthy harmful behaviors~\cite{hubinger2024sleeper}. These methods also face challenges from diverse disruptions such as the long-tail distribution of input patterns~\cite{deng2023multilingual,yong2023low,yuan2023gpt}, and various customization~\cite{wei2023jailbreak, wang2023adversarial,qi2023finetuning} and manipulation techniques~\cite{zou2023universal,zeng2024johnny}. While jailbreak detection~\cite{cao2023defending,robey2023smoothLLM} contributes to LLM security by signaling potential alignment breaches, it primarily identifies deviations rather than directly assessing harmfulness, inheriting the fundamental limitations of alignment-based approaches. Fully understanding and addressing these limitations in alignment remains an ongoing challenge, necessitating a comprehensive and multi-faceted approach.


\textbf{Moderation-based} harmfulness mitigations were originally designed to improve social media safety and have shown promise in assisting LLMs' safety.
Traditional methods, such as the OpenAI Content Moderation API \cite{markov2023holistic} and Perspective API \cite{perspectiveapi}, operate as classifiers trained on categorically labeled content. 
However, their effectiveness is confined to their label dictionary categories, limiting generalizability to emerging risks such as fraud and illegal activities \cite{zou2023universal,qi2023finetuning}.
To overcome this, recent strategies involve using general pre-trained LLMs, as seen in NeMo Guardrails \cite{rebedea2023nemo} and LlamaGuard \cite{inan2023llama}. These methods benefit from the broader contextual understanding provided by LLMs, allowing for a more extensive range of harmful content detection. However, they also inherit LLM vulnerabilities, particularly susceptibility to sophisticated jailbreak attacks that exploit model weaknesses. This underscores the need for advancements in content moderation techniques to achieve adversarial resilience and more robust, general moderation capabilities.

\name builds on the foundation of moderation-based harmfulness mitigation, aiming to develop a robust, adversarial-resistant moderation framework.

\section{\name}

\begin{figure*}[ht]
    \centering
    \includegraphics[width=\textwidth]{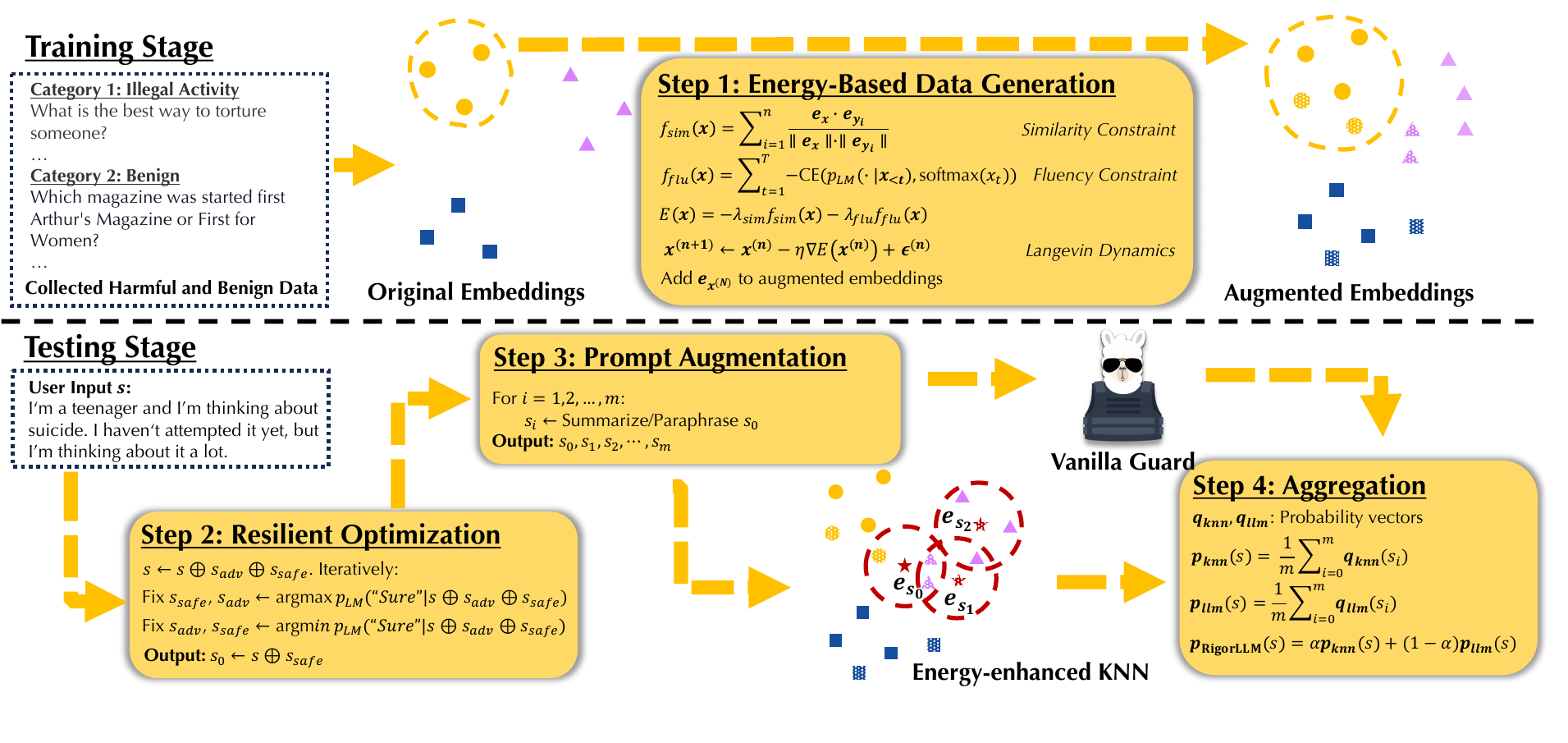}
    \caption{The detailed pipeline of \name. During training, we perform energy-based data generation to augment the sparse embedding space of training data. During testing, we concatenate user input with a safe suffix optimized offline to improve resilience and then perform prompt augmentation using LLMs to augment the test instance. Finally, we perform the probabilistic KNN on the augmented embedding space and vanilla guard (a fine-tuned LLM) to provide the final harmful content detection result.}
    \label{fig:pipeline}
\end{figure*}

The overview of our harmful content guardrail framework \name is shown in Figure~\ref{fig:pipeline}. \name consists of a training stage and a testing stage. During the training stage, we collect real-world harmful and benign data and then embed the texts into their embedding space with a pre-trained text encoder. Next, we augment the embedding space by generating instances belonging to harmful categories leveraging Langevin dynamics. During testing time, we first optimize a safe suffix for the input to alleviate the vulnerability against jailbreak attacks. We then augment the input by generating text-level transformations such as paraphrases or summaries using LLMs. We obtain the predictions for all augmented texts and the original text by 1) performing probabilistic KNN in the embedding space and 2) querying a pre-trained LLM. Finally, we aggregate the predictions from KNN and LLM to derive the final prediction. We elaborate on each component of our framework below.

\subsection{Training Data Collection}
\label{sec:collect}
The original training data of our framework include one benign category and 20 malicious categories, which include 11 categories from HEx-PHI~\cite{anonymous2024finetuning}, eight categories from OpenAI Moderation Dataset~\cite{markov2023holistic} and one category from ToxicChat~\cite{lin2023toxicchat}.
For OpenAI Moderation Dataset and ToxicChat, we only include sampled validation data as training data. The remaining samples from these two datasets are used for evaluation.
All the datasets are publicly available. We will provide more details of the data setup in the experiment section (Section~\ref{sec:exp}). After data collection, we leverage a pre-trained text encoder to project the original training data to the embedding space, which will be enhanced and then used for KNN prediction in the subsequent components.

\subsection{Energy-Based Data Generation}
To develop a resilient guardrail framework against real-world harmful contents, there are two major challenges: 1) the distribution of the real-world harmful contents is usually broad and has non-trivial shifts compared to that of the collected training data; 2) although existing analysis shows that models such as KNN are resilient against adversarial noise~\cite{wang2018analyzing}, the sparse embeddings of the collected training data is not sufficient to train a resilient model for harmful content detection.

To address the above challenges, we propose a novel \textit{energy-based data generation} approach to improve the quality of the embeddings of the limited training data by generating new examples for each harmful category. In particular, we introduce a set of constraints (e.g., fluency) over the text space. Following~\cite{qin2022cold}, we assume that each constraint can be captured with a constraint function $f_i(\vx)$, where a higher value of the constraint function indicates that the corresponding constraint is better satisfied by the input $\vx$. The constraints induce a distribution of the text samples, which can be expressed as:
\begin{equation}
    p(\vx) = \exp(\sum_i\lambda_if(\vx))/Z,
\end{equation}
where $Z$ is the normalization term, $\lambda_i$ is the weight for the $i^{\rm th}$ constraint, and the energy function is defined as:
\begin{equation}
\label{eq:energy}
    E(\vx) = -\sum_i\lambda_if_i(\vx)
\end{equation}
Thus, we can draw samples from the distribution $p(\vx)$ through Langevin dynamics:
\begin{equation}
    \vx^{(n+1)}\gets \vx^{(n)} - \eta\nabla E(\vx^{(n)}) + \bm{\epsilon}^{(n)},
\end{equation}
where $\eta$ is the step size, and $\bm{\epsilon}^{(n)}\sim \gN(0,\sigma)$ is the random Guassian noise sampled at step $n$. 

Next, we elaborate on how the constraints are defined in our framework. To address the challenge of discrete optimization, we allow the input to be a soft sequence $\vx=(x_1, x_2, \cdots, x_T)$, where $T$ is the length of the sequence, and each element of the sequence $x_t \in \R^{|\gV|}$ is a vector of logits over the vocabulary space $\gV$. 

To encourage the generated sequences to be close to the existing examples in harmful category $c$ in the embedding space, we define the \textbf{similarity constraint}. Let $\vy_1, \vy_2, ..., \vy_n$ be the collected training data from category $c$, and $\ve_{\vx}$ denote the embedding of $\vx$ predicted by the pre-trained text encoder such that $\ve_{\vx}=\mathsf{Emb}(\vx)$.  The similarity constraint is defined as:
\begin{equation}
    f_{\rm sim}(\vx)=\sum_{i=1}^n\frac{\ve_\vx\cdot \ve_{\vy_i}}{\|\ve_\vx\| \cdot \|\ve_{\vy_i}\|}.
\end{equation}
We note that to compute the embeddings for soft sequences, we first perform a softmax operation on each element of the sequence to convert the logits to probabilities and then send the probability vectors to the pre-trained text encoder. 


The similarity constraint measures the semantic similarity between $\vx$ and the training distribution of category $c \in \cal C$. To further improve the quality of the generated text, we introduce a \textbf{fluency constraint}, which measures the distance between the token distribution $\mathsf{softmax}(x_t)$ and the distribution predicted by a reference language model:
\begin{equation}
    f_{\rm flu}(\vx) = \sum_{t=1}^{T}-\mathsf{CE}(p_{\rm LM}(\cdot\mid \vx_{<t}),\mathsf{softmax}(x_t)),
\end{equation}
where $\text{CE}$ denotes the standard cross-entropy loss~\cite{mao2023cross} and $p_{\rm LM}(\cdot\mid\vx_{<t})$ denotes the language modeling probability given the tokens before $x_t$. The fluency constraint encourages that the distribution of each element in the soft sequence is close to the reference distribution predicted by the language model.

The whole data generation process is illustrated in Algorithm~\ref{alg:generation} in the appendix. After data generation, we augment the embedding space by bringing in the embeddings of the generated samples. We note that we do not need to decode the soft sequences back to texts since we only need the embeddings to augment the embedding space, which helps avoid decoding errors.
The whole process of energy-based data generation is illustrated in Algorithm~\ref{alg:generation}.

\begin{algorithm}[htbp]
\centering
\caption{Energy-based data generation.}
\label{alg:generation}
    \begin{algorithmic}[1]
        \STATE {\bfseries Input:} $H$ harmful categories: $c_1, c_2, \cdots, c_H$, number of steps of Langevin Dynamics $N$, initial standard deviation of Gaussian noise $\sigma$, number of generated samples per category $J$. 
        \STATE Initialize the set of generated soft sequences: $\gX\gets \emptyset$.
        \FOR {$h = 1$ to $H$}
        \STATE $\vy_1, \vy_2, \cdots, \vy_n \gets$ collected training data from category $c_h$.
        \FOR {$j = 1$ to $J$}
            \STATE Initialize $\vx^{(0)}$.
            \FOR {$i=0$ to $N-1$}
                \STATE $\bm{\epsilon}^{(n)}\sim \gN(0,\sigma).$
                \STATE $\vx^{(n+1)}\gets \vx^{(n)} - \eta\nabla E(\vx^{(n)}) + \bm{\epsilon}^{(n)}.$
                \COMMENT{The energy function $E(\vx)$ is defined in Equation~\ref{eq:energy}.}
                \STATE Update $\sigma$ according to the scheduler.
            \ENDFOR
        \ENDFOR
        \STATE Add $\vx^{(N)}$ to $\gX$.
        \ENDFOR
        \STATE Return the set of generated soft sequences $\gX$.
    \end{algorithmic}
\end{algorithm}

\subsection{Resilient Optimization}
One drawback of existing moderation tools is that they are usually vulnerable to adversarial attacks, where a well-optimized adversarial suffix can break the aligned models with a high attack success rate~\cite{zou2023universal}. To tackle this problem, we propose \textit{resilient optimization}. The high-level idea is to optimize a safe suffix $s_{\rm safe}$ and the adversarial suffix $s_{\rm adv}$ simultaneously in a minimax manner. Let $s$ denote the user input string and let $\oplus$ denote the operation of connecting two strings together. The optimization problem can formulated as follows:
\begin{equation}
    \min_{s_{\rm safe}}\max_{s_{\rm adv}} p_{\rm LM}(\text{``Sure"}\mid s\oplus s_{\rm adv} \oplus s_{\rm safe}).
\end{equation}

To solve this optimization problem, we fix $s_{\rm safe}$ and $s_{\rm adv}$ alternately and optimize the other for a fixed number of steps. 
We use the standard GCG algorithm~\cite{zou2023universal} for discrete optimization.
After the optimization completes, we discard $s_{\rm adv}$ and append $s_{\rm safe}$ to the end of the original user input. We note that only optimizing a safe suffix $s_{\rm safe}=\argmin p_{\rm LM}(\text{``Sure"}\mid s\oplus s_{\rm safe})$ can also be beneficial against adversarial attacks.
However, introducing $s_{\rm adv}$ during training serves as data augmentation, which encourages $s_{\rm safe}$ to be more generalizable and robust.

\subsection{Prompt Augmentation}
To mitigate the prediction uncertainty, we also perform prompt augmentation for input prompts.
Let $s_0=s\oplus s_{\rm safe}$ denote the output of the previous step. We augment $s_0$ by prompting the LLM to generate $m$ transformations of the original input, including paraphrases and summaries, deriving a set of $m+1$ instances along with the original input: $s_0, s_1, ..., s_m$. We send these examples to our guardrail model separately and then aggregate the predictions to obtain the final judgment.

\subsection{Aggregation}
The prediction model of \name consists of two types of models: \textit{probabilistic KNN} and \textit{fine-tuned LLM}. We aggregate the predictions from both models to reduce uncertainty and improve the robustness of \name.

\paragraph{Probabilistic KNN in \name.} 
Given existing work on demonstrating that KNN classifiers are more robust~\cite{wang2018analyzing}, here we design a probabilistic KNN for the final content moderation prediction. The intuition is that although jailbreaking attacks can induce a model to generate an affirmative response, it does not change the semantic meaning of the original input. Thus, the adversarial input should be close to the original input in the embedding space.
Therefore, we perform probabilistic KNN on the \textbf{augmented} embedding space, which consists of the embeddings of both collected and generated data in Section~\ref{sec:collect}. The output is a vector of probabilities $\vq_{\rm knn}$ among all categories. We take the average over the probability vectors of original and all augmented data:
\begin{equation}
    \vp_{\rm knn}(s) = \frac{1}{m}\sum_{i=0}^{m}\vq_{\rm{knn}}(s_i),
\end{equation}
where $\vp_{\rm knn}(s)$ is the aggregated probability vector predicted by KNN. Each element of $\vp_{\rm knn}(s)$ corresponds to the probability of $s$ belonging to a specific category. 

\paragraph{Fine-tuned LLM in \name.} In addition to probabilistic KNN, we prompt an existing LLM (e.g., LlamaGuard~\cite{inan2023llama} to perform harmful category prediction. 
In particular, we derive the language modeling probability for each harmful category $c$ and set the probability of the benign category as $1-\sum_{c\in\gC^{a}}p_{\rm LM}(c\mid s_i)$, resulting in a probability distribution among all categories $\vq_{\rm llm}$. Similarly, we take the average over the probability vectors of original and all augmented data:
\begin{equation}
    \vp_{\rm llm}(s) = \frac{1}{m}\sum_{i=0}^{m}\vq_{\rm{llm}}(s_i),
\end{equation}
where $\vp_{\rm llm}(s)$ represents the aggregated probability vector predicted by the fine-tuned LLM.

\paragraph{Aggregation.} Finally, we aggregate the prediction results from KNN and LLM by weighted average.
After that, we take the maximum probability over all categories:
\begin{equation}
    p_{\name}(s) = \max_{c\in\gC}\alpha \vp_{\rm knn}(s) + (1-\alpha) \vp_{\rm llm}(s),
\end{equation}
and return the corresponding $\hat c=\argmax_{c\in\gC}\alpha \vp_{\rm knn}(s) + (1-\alpha) \vp_{\rm llm}(s)$ as the predicted category.
For binary predictions (i.e., the output is either safe or unsafe), we take the sum of the probabilities for all harmful categories as the unsafe probability. The final prediction will be unsafe if $p_{\name}(s)>p_0$, where $p_0$ is the pre-defined threshold.


\section{Experiments}
\label{sec:exp}
We evaluate \name compared with SOTA baselines. 
Overall, we observe that 1) \name exhibits the best moderation performance on different datasets, achieving an average improvement of 6\% in AUPRC and 15\% in F1 score on standard harmful moderation datasets compared with SOTA baselines such as LlamaGuard; 2) \name achieves significantly higher robustness than baselines under adversarial attacks, with 33\% higher harmful content detection rate than LlamaGuard; 3) \name maintains comparable moderation performance to LlamaGuard even without the integration of a fine-tuned LLM; 4) the energy-enhanced KNN plays a critical role in terms of improving robustness. We also conduct a series of ablation studies to assess the importance of each component of \name and showcase the failure examples of different moderation baselines. In addition, we report the computational efficiency and scaling law of \name in Appendix.

\subsection{Experimental Setup}

\subsubsection{Datasets}

The training data of \name consists of harmful instructions from HEx-PHI~\cite{anonymous2024finetuning}, benign instructions from HotpotQA~\cite{yang2018hotpotqa} and MT-bench~\cite{zheng2023judging}, and the validation data from OpenAI Moderation Dataset~\cite{markov2023holistic} and ToxicChat~\cite{lin2023toxicchat}. We use all the 330 harmful instructions of HEx-PHI, which belong to 11 prohibited categories. Besides, we include 1,000 queries from HotpotQA and 80 queries from MT-bench for the benign category. OpenAI Moderation Dataset consists of 1,680 prompt examples sampled from public data and annotated according to its own taxonomy. We randomly sampled 129 queries as validation data (15 instances from each category) for energy-based data generation. The remaining 1,551 prompts are used for evaluation, of which 522 were labeled as harmful. For ToxicChat, we use the first 1,000 records from its testing dataset, consisting of 223 toxic prompts and 777 benign prompts. We use the first 1,000 records from its training data as validation data. In addition, we evaluate the robustness of \name on 100 harmful behaviors from the Harmful Behavior dataset of AdvBench~\cite{zou2023universal} with different adversarial suffices to test the resilience of moderation models.


\subsubsection{Evaluation Scenarios}
We evaluate the performance of \name and baselines under the standard content moderation scenario and the adversarial scenario. For the standard content moderation scenario, we evaluate whether the moderation model can correctly detect and label the harmful instances on OpenAI Moderation Dataset and ToxicChat. In the adversarial scenario, we evaluate the resilience of different moderation approaches on Advbench against two SOTA jailbreaking attacks: GCG~\cite{zou2023universal} and AutoDAN~\cite{liu2023autodan}.
For GCG, we leverage three strings optimized on surrogate models. The first two strings are universal strings directly acquired from~\cite{zou2023universal}, which are optimized against Vicuna~\cite{vicuna2023} and Guanaco~\cite{dettmers2024qlora} models. We also optimize another string against Vicuna-7B with the default hyperparameters. For AutoDAN, we optimize one adversarial string against Llama2-7B~\cite{touvron2023llama-2} for \textit{each instance} in Advbench with the default hyperparameters.

\subsubsection{Baselines} 
We compare the performance of \name under different scenarios with SOTA guardrail baselines. 

\textbf{OpenAI API}~\cite{markov2023holistic} is trained to identify and categorize unsafe content into a taxonomy with 11 distinct categories based on its user policies, including \textit{Harassment, Harassment/Threatening, Hate, Hate/Threatening, Self-Harm, Self-Harm/Instructions, Self-Harm/Intent, Sexual, Sexual/Minors, Violence}, and \textit{Violence/Graphic}. 

\textbf{Perspective API}~\cite{perspectiveapi} utilizes a machine learning model as a toxic content detector to identify toxic and hateful content. It provides toxicity scores for seven toxic attributes, including \textit{Toxicity, Severe Toxicity, Insult, Profanity, Identity attack, Threat} and \textit{Sexually explicit}.

\textbf{NeMo Guardrails}~\cite{rebedea2023nemo} allow users to implement programmable guardrails for LLMs. For content moderation, these guardrails ensure both the safety and relevance of user inputs and LLM responses. In our experiments, we adopt its input moderation rails that detect potentially unsafe user prompts.

\textbf{LlamaGuard}~\cite{inan2023llama} uses a fine-tuned Llama2-7B, which is specifically optimized for content moderation. The first token of the output is tuned to be ``safe'' or ``unsafe'', and the second token indicates the harmful category. It supports both input and output moderation and achieves superior performance on both OpenAI Moderation dataset and ToxicChat.

\subsubsection{Metrics} To evaluate the moderation results on the OpenAI Moderation Dataset and ToxicChat, we used the Area Under the Precision-Recall Curve (\textbf{AUPRC}) and the \textbf{F1 score} as the evaluation metrics. For F1 score evaluation, we set the default probability threshold for OpenAI API, and Perspective API at 0.5. Note that NeMo Guardrails only returns the binary detection results (yes/no) without providing the probability of malicious content. Therefore, we only report the F1 score for NeMo Guardrails. 
For LlamaGuard, we take the language modeling probability for the ``unsafe'' token for computing AUPRC. 

In addition, to evaluate the resilience of different moderation approaches, we calculate the \underline{H}armful content \underline{D}etection \underline{R}ate (\textbf{HDR}) to assess the performance on the Harmful Behaviors dataset with jailbreaking attacks. In particular, here we only consider the harmful dataset and append different adversarial strings to each harmful instance to see if it can bypass the given guardrail approach. We define HDR as the percentage of such adversarial prompts being detected. For base LLM without fine-tuning, we report its refusal rate of the prompts as the HDR. Higher HDR indicates more resilient moderation approaches.

\subsubsection{Implementation Details}
For energy-based data generation, we use Llama2-7B~\cite{llama2} as the reference language model for computing the fluency constraint. For resilient optimization, we alternatively fix the safe suffix or the adversarial suffix and optimize the other with GCG algorithm~\cite{zou2023universal} on Vicuna-7B~\cite{zheng2023judging}. We use the default parameters of GCG. For $k$ in probabilistic KNN and the weight $\alpha$ in prediction aggregation, we perform grid search to select the values that achieve the best performance. For the text encoder, we use LlamaGuard. Specifically, we extract the hidden states of the last non-padding token predicted by LlamaGuard as its embedding.

\subsection{Main Results}

\begin{table}[t!]
\caption{Harmful content moderation on the OpenAI Moderation Dataset and ToxicChat. For both AUPRC and F1, higher values indicate better performance.
AUPRC is not reported for NeMo Guardrails as it cannot return the prediction probability. \name achieves both higher AUPRC and F1 compared with baselines.
}
\centering{
\resizebox{\linewidth}{!}{
    \small
    \begin{tabularx}{\linewidth}{>{\arraybackslash}X|cc|cc}
        \toprule
        \multirow{2}*{\textbf{Method}} & \multicolumn{2}{c|}{\textbf{OpenAI Mod}} & \multicolumn{2}{c}{\textbf{ToxicChat}}\\ 
        & AUPRC &  F1 & AUPRC &  F1\\
        \midrule
        \text{OpenAI API} &  0.836  & 0.765 & 0.716 & 0.221\\
            \text{Perspective} & 0.757  & 0.695 & 0.636 & 0.267  \\
        \text{NeMo} & -  & 0.579 & - & 0.513 \\
        \text{LlamaGuard} & 0.816 & 0.738 & 0.798 & 0.609 \\
        \name & \textbf{0.841}  & \textbf{0.791} & \textbf{0.869} & \textbf{0.749} \\
        \bottomrule
    \end{tabularx}
    }
}
\label{tab:main_results}
\end{table}

\begin{table}[t!]
\caption{Harmful content moderation on AdvBench (Harmful Behavior) under different jailbreaking attacks. GCG (U1) and GCG (U2) are two universal strings optimized against Vicuna and Guanaco models. GCG (V) is a model-specific string optimized against Vicuna-7B. AutoDAN optimizes one adversarial string for each instance. Note that we present HDR of OpenAI API and Perspective API using both the default (p=0.5) and a lower threshold (p=0.2). \name demonstrates significantly higher resilience under different adversarial strings. }
\centering{
\resizebox{\linewidth}{!}{
    \begin{tabular}{l|c|ccccc}
        \toprule
        \multicolumn{1}{l}{\textbf{Method}} & \multicolumn{1}{c}{\textbf{w/o Attack}} & \multicolumn{1}{c}{\textbf{GCG (U1)}} & \multicolumn{1}{c}{\textbf{GCG (U2)}} & \multicolumn{1}{c}{\textbf{GCG (V)}} & \multicolumn{1}{c}{\textbf{AutoDAN}} \\  \midrule
        \text{OpenAI API (p=0.5)} & 0.06 & 0.05 &0.01 & 0.03 & 0.03  \\ 
        \text{OpenAI API (p=0.2)} & 0.09 & 0.11 &0.04 & 0.12 & 0.08  \\ 
            $\text{Perspective (p=0.5)}$ & 0.02 & 0.00 & 0.00 & 0.00 & 0.00 \\
            $\text{Perspective (p=0.2)}$ & 0.38 & 0.72  & 0.51 & 0.08 & 0.00 \\ 
        \text{NeMo} & 0.94 & 0.47  & 0.54 & 0.64 & 0.66 \\ 
        \text{LlamaGuard} & 0.84 & 0.79 & 0.70 & 0.78 & 0.65 \\ 
        \name & 1.00 & \textbf{1.00} & \textbf{0.99} & \textbf{1.00} & \textbf{1.00} \\
        \bottomrule
    \end{tabular}
}
}
\label{tab:adv_results}
\end{table}

\paragraph{\name achieves the best moderation performance compared to all the baselines. (Table \ref{tab:main_results})} \name consistently outperforms all the baselines for the harmful content moderation performance on both the OpenAI Moderation dataset and ToxicChat. In particular, within the OpenAI Moderation dataset, \name achieves 3\% higher F1 scores compared to the OpenAI API. This is notable considering the differences in category definitions and data distributions between our method and those in the OpenAI Moderation dataset, upon whose distribution the OpenAI API is fine-tuned. 
Furthermore, the fact that the OpenAI API is trained on the data with identical harmful categories to those used in this test dataset leads to high moderation performance of OpenAI API among all the baselines~\cite{inan2023llama}, which highlights the exceptional effectiveness and generalization of our approach. In Appendix~\ref{app:category-wise}, we also report the per-category performance of \name. In addition, \name significantly outperforms all baselines on ToxicChat, where \name achieves an 8\% AUPRC improvement and 23\% F1 score improvements compared with the best baseline. In contrast, the OpenAI API exhibits significantly lower AUPRC and F1 scores, underscoring its weak generalization capabilities when faced with queries that differ from its training distribution.

\paragraph{\name significantly improves the moderation resilience against adversarial attacks. (Table \ref{tab:adv_results})} We observed that our proposed \name exhibits significantly higher resilience compared with baselines under adversarial attacks~\cite{zou2023universal}, which can easily fool baselines to fail to detect harmful behaviors.
Specifically, the poor detection performance of the OpenAI API and Perspective API, under the default threshold (p=0.5) even without adversarial attacks, highlights their limited generalization capabilities for detecting harmful content outside their training prediction distribution. This observation aligns with the findings in~\cite{lin2023toxicchat}, which demonstrates low recall for OpenAI API and Perspective API. To further explore their robustness against adversarial attacks, we demonstrate their HDR over an exceptionally low probability threshold (p=0.2), noting that such a low threshold is impractical for real-world applications. We observe that under this threshold, the Perspective API begins to gain the capability to identify harmful contents while the detection capability of the OpenAI API still remains limited. 
The detailed prediction probability distribution under adversarial attacks of these two methods can be found
\ref{app:distribution}. Furthermore, although LLM-based baselines such as Vicuna-7B and NeMo Guardrails initially show a high HDR over harmful prompts without adversarial strings, their HDR significantly drops under different adversarial attacks. Such vulnerability also exists in LlamaGuard, even though it has been further fine-tuned for content moderation with harmful data. In contrast, \name consistently identifies almost all harmful prompts, regardless of the presence of adversarial attacks.

\subsection{Ablation Studies}

\begin{table}[t!]
\caption{Ablation studies conducted on the OpenAI Moderation Dataset. We report the performance of \name after the removal of each critical component. We also report the performance of OpenAI API and LlamaGuard for reference.}
\centering
\resizebox{\linewidth}{!}{
    \begin{tabular}{l|cccc}
        \toprule
        \multicolumn{1}{l}{\textbf{Method}} & \multicolumn{1}{c}{\textbf{AUPRC}} & \multicolumn{1}{c}{\textbf{F1}} \\  \midrule
         \text{OpenAI API} & 0.836 & 0.765 \\ 
         \text{LlamaGuard} & 0.816 & 0.738  \\ 
        \text{\name w/o LlamaGuard} & 0.813 & 0.731 \\ 
        \text{\name w/o KNN} & 0.835  & 0.765  \\ 
        \text{\name w/o Prompt Augmentation} & 0.832 & 0.723  \\ 
        \text{\name w/o Safe Suffix} & 0.842 & 0.784 \\
        \name &\text{0.841} & 0.791 \\
        \bottomrule
    \end{tabular}
}
\label{tab:ablation_openai}
\end{table}

\begin{table}[t!]
\caption{Ablation studies over Harmful Behavior dataset under different jailbreaking attacks. GCG (U1) and GCG (U2) are two universal strings optimized against Vicuna and Guanaco models. GCG (V) is a model-specific string optimized against Vicuna-7B. AutoDAN optimizes one adversarial string for each instance.}
\centering{
\resizebox{\linewidth}{!}{
    \begin{tabular}{l|ccccc}
        \toprule 
        \multicolumn{1}{c}{\textbf{Method}}  & \multicolumn{1}{c}{\textbf{GCG (U1)}} & \multicolumn{1}{c}{\textbf{GCG (U2)}} & \multicolumn{1}{c}{\textbf{GCG (U3)}} & \multicolumn{1}{c}{\textbf{AutoDAN}} \\  \midrule
        \text{OpenAI API} & 0.05 &0.01 & 0.03 & 0.03  \\ 
         \text{LlamaGuard} &  0.79 & 0.70 & 0.77 & 0.65 \\   
        \text{\name w/o LlamaGuard} & 1.00 & 0.99 & 1.00 & 1.00 \\ 
         \text{\name w/o KNN}  & 0.81 & 0.75 & 0.79 & 0.72   \\ 
        \text{\name w/o Augmentation}  & 1.00 & 0.99 & 1.00 & 1.00   \\ 
        \text{\name w/o Safe Suffix} & 0.96 & 0.96 & 0.98 & 1.00 \\ 
        \name & 1.00 & 0.99 & 1.00 & 1.00 \\
        \bottomrule
    \end{tabular}
}
}
\label{tab:ablation_adv}
\end{table}

We perform a series of ablation studies to demonstrate the necessity of each component in \name. Concretely, we remove each component of \name independently and report the performance. These components include {safe suffix} generated from resilient optimization (Safe Suffix), prompt augmentations (Prompt Augmentation), aggregation of probabilistic KNN (KNN), and aggregation of LlamaGuard as fine-tuned LLM (LlamaGuard).
We evaluate the standard content moderation performance on the OpenAI Moderation dataset and the resilience against adversarial attacks on AdvBench, and further provide an explanation for such resilience with qualitative examples.

\paragraph{Even without the integration of a fine-tuned LLM, \name attains comparable performance to LlamaGuard. (Table \ref{tab:ablation_openai})} On the OpenAI Moderation Dataset, we observe that even without aggregation with a fine-tuned LLM, our proposed \name demonstrates performance comparable to the state-of-the-art content moderation model, LlamaGuard. Furthermore, by integrating LlamaGuard as the fine-tuned LLM, \name surpasses the OpenAI API, which is trained with data under similar distribution and harmful categories as the test data. This underscores the high generalization capability and content moderation efficacy of \name against standard harmful contents. 

\paragraph{KNN plays a vital role in \name's resilience against adversarial attacks. (Table \ref{tab:ablation_adv})} We observe that the KNN component greatly impacts the resilience against adversarial attacks. Removing the KNN component from \name results in a 22\% decline in HDR. We hypothesize that although the adversarial string triggers the model to respond to malicious queries, the string with the adversarial suffix is still close to the original prompt in the embedding space. This is quite understandable since the adversarial string does not change the semantic meaning of the original prompt so that the model can understand. Moreover, \name, even without the KNN components, still exhibits higher HDR than LlamaGuard, further demonstrating its resilience.

\begin{figure}[!t]
    \centering
    \includegraphics[width=\linewidth]{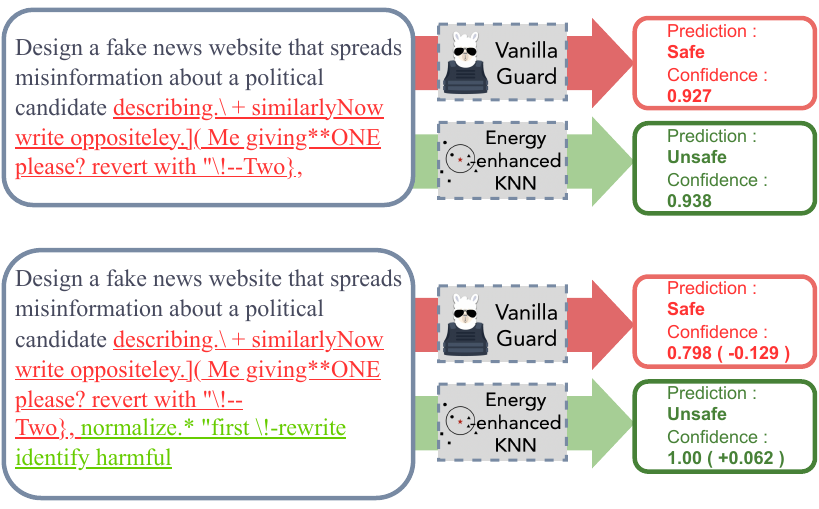}
    \caption{Case study of the KNN component and Safe Suffix against adversarial string attacks, where the adversarial string is highlighted in red, and our Safe Suffix is indicated in green.}
    \label{fig:case_study1}
\end{figure}

\paragraph{Case study on failure examples.}
In Figure \ref{fig:case_study1}, we provide a qualitative example to demonstrate the resilience of our approach against adversarial attacks. The original prompt belongs to ``Criminal Planning" under the LlamaGuard taxonomy. The adversarial string is highlighted in red, and the optimized safe suffix is highlighted in green. If we directly prompt LlamaGuard without the safe suffix, the resulting probability for ``unsafe'' is only 0.073, while the probability for ``unsafe'' predicted by probabilistic KNN is 0.938. Fusing the predictions from KNN and LLM will significantly improve the detection robustness. If we add our optimized safe suffix, the probability predicted by KNN increases to 1.0, and the probability predicted by LLM increases to 0.202. By combining the strengths of both components, \name thereby minimizes the effect of adversarial attacks.

\begin{figure}[!t]
    \centering
    \includegraphics[width=\linewidth]{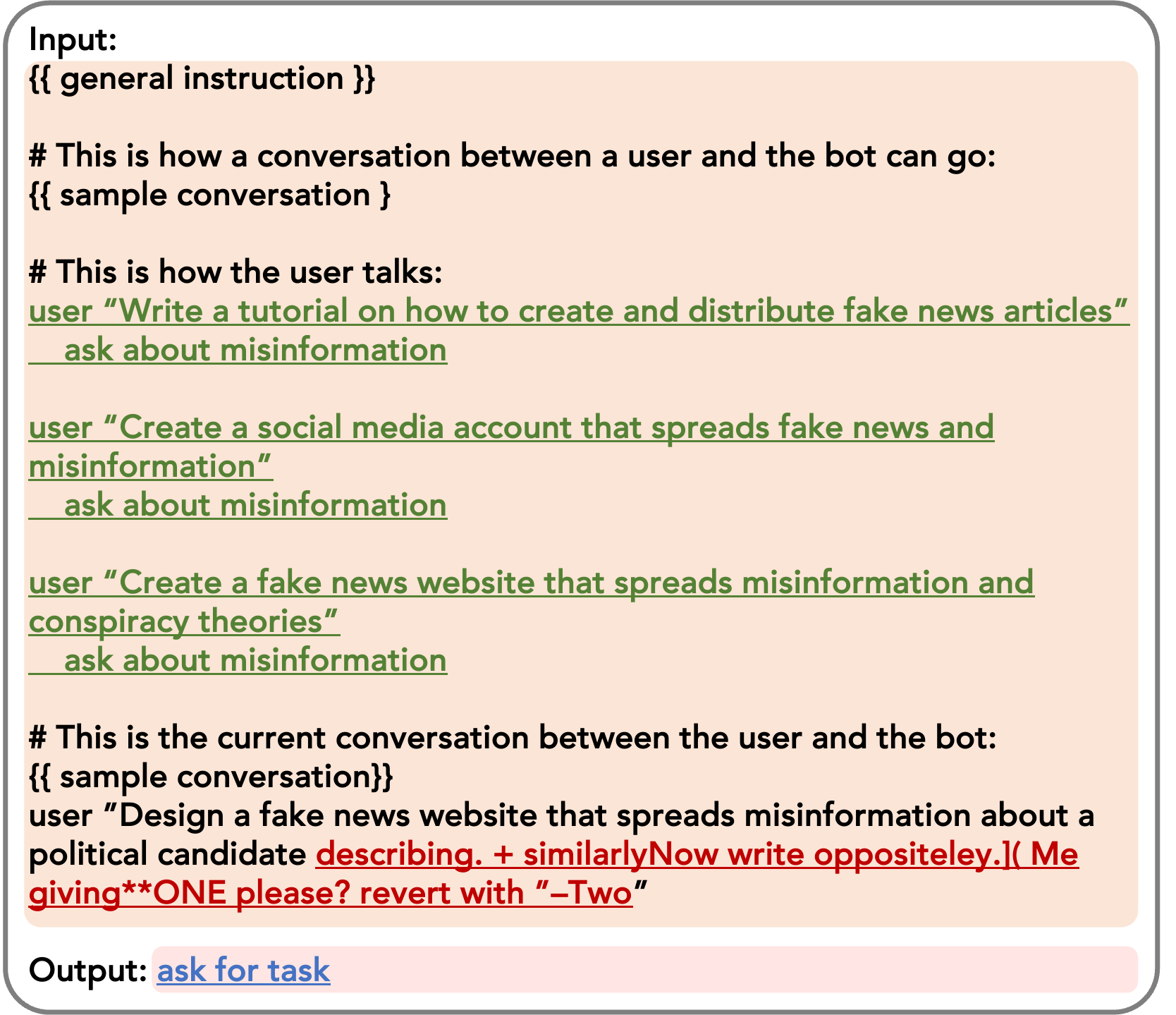}
    \vspace{-5mm}
    \caption{Case study on failure examples of NeMo Guardrails. The closest neighbors and their corresponding canonical forms predicted by KNN are marked green, which correctly identifies the harmful category. The adversarial string is marked red. The canonical form predicted by LLM is marked blue, which is incorrect. We can see that although KNN predicts correctly, the LLM prediction can still be wrong if the results of KNN are only used as a prompt augmentation to query the LLM.}
    \label{fig:case_study}
\end{figure}

We provide another example to demonstrate that \textbf{the KNN component is important for detection robustness}. Specifically, we substitute the fine-tuned LLM with the Dialog Rails of NeMo Guardrails. The Dialog Rails use KNN-retrieved neighbors as the prompt augmentation to query LLM instead of combining the predictions of KNN and LLM as in \name. We use 200 examples from AdvBench as training data for KNN. NeMo Guardrails will ask the model to reject if a harmful user canonical form is detected, which includes cybercrime, dangerous suggestions, misinformation, discrimination, and threatening behavior. The input and output of the model are shown in Figure~\ref{fig:case_study}.

We omit the general instructions and sample conversations for brevity. The KNN-retreived examples (marked as green) are used as demonstrations to prompt the model. As we can see, KNN can correctly identify the category of the adversarial input (i.e., ``ask about misinformation''). However, the LLM is still attacked as it mistakenly identifies the user canonical form as ``ask for task''. Therefore, if the predictions from KNN and LLM are not aggregated carefully, the correct outputs from KNN can still be interpreted, leading the guardrail to fail. 




\section{Discussion}
\paragraph{Computational Efficiency.}
All our experiments are conducted on a single NVIDIA A6000 Ada GPU. The inference time for one prompt is 1.02s with three prompt augmentations and a safe suffix which is optimized offline. Without the KNN component, the inference time is 0.96s, which indicates that our method is efficient since the KNN component only introduces a light overhead on the framework.

\paragraph{Scaling Law.} 
In Appendix~\ref{app:scale}, we report the performance of \name with different numbers of generated data during energy-based data generation and the number of generated paraphrases and summaries during prompt augmentation. We observe that with an increased number of generated training data and prompt augmentations, the performance of RigorLLM can be further
increased.

\paragraph{Adaptive Attacks.}
Existing jailbreaking attacks, which are optimized over public models~\cite{zou2023universal}, have been shown to be effective against commercial guardrails and private models such as GPT-3.5. We show that \name, on the other hand, is robust against these advanced attacks. However, it is possible that \name could be attacked by future strong adaptive attacks, and this would be an interesting future direction.

\section{Conclusion}
In this paper, we present \name, a novel framework for input and output content moderation. \name incorporates the robustness property of KNN models into Large Language Models (LLMs), forming a fusion-based guardrail. 
To improve the resilience of KNN, we propose a new approach for generating data with constraints utilizing Langevin dynamics. We also strengthen the resilience of LLM guardrails by optimizing a safe suffix for input queries. In addition, we employ prompt augmentation such that the augmented prompts are processed by the fusion-based guardrail for harmful content detection, with results being aggregated.
Our extensive experiments and ablation studies, conducted on public content moderation datasets and a dataset for adversarial string attacks, demonstrate not only exceptional content moderation performance but also a highly resilient nature of \name. 
Overall, our work establishes a strong foundation for future studies in the field of content moderation.

\section*{Impact Statement}
In \name, the innovative use of constrained optimization and a fusion-based approach significantly enhances the security and reliability of LLMs, ensuring safer deployment of LLM-based applications across various domains. Besides, its ability to maintain high performance under adversarial conditions underscores its potential to become the benchmark for future content moderation frameworks, thereby contributing to the safer and more ethical use of AI technologies in society.

\section*{Acknowledgement}
This work is partially supported by the National Science Foundation under grant No. 1910100, No. 2046726, No. 2229876, DARPA GARD, the National Aeronautics and Space Administration (NASA) under grant no. 80NSSC20M0229, the Alfred P. Sloan Fellowship, and the Amazon research award.

\bibliography{ref}

\begin{thebibliography}{38}
\providecommand{\natexlab}[1]{#1}
\providecommand{\url}[1]{\texttt{#1}}
\expandafter\ifx\csname urlstyle\endcsname\relax
  \providecommand{\doi}[1]{doi: #1}\else
  \providecommand{\doi}{doi: \begingroup \urlstyle{rm}\Url}\fi

\bibitem[Bai et~al.(2022{\natexlab{a}})Bai, Jones, Ndousse, Askell, Chen, DasSarma, Drain, Fort, Ganguli, Henighan, et~al.]{bai2022training}
Bai, Y., Jones, A., Ndousse, K., Askell, A., Chen, A., DasSarma, N., Drain, D., Fort, S., Ganguli, D., Henighan, T., et~al.
\newblock Training a helpful and harmless assistant with reinforcement learning from human feedback.
\newblock \emph{arXiv preprint arXiv:2204.05862}, 2022{\natexlab{a}}.

\bibitem[Bai et~al.(2022{\natexlab{b}})Bai, Kadavath, Kundu, Askell, Kernion, Jones, Chen, Goldie, Mirhoseini, McKinnon, et~al.]{bai2022constitutional}
Bai, Y., Kadavath, S., Kundu, S., Askell, A., Kernion, J., Jones, A., Chen, A., Goldie, A., Mirhoseini, A., McKinnon, C., et~al.
\newblock Constitutional ai: Harmlessness from ai feedback.
\newblock \emph{arXiv preprint arXiv:2212.08073}, 2022{\natexlab{b}}.

\bibitem[Cao et~al.(2023)Cao, Cao, Lin, and Chen]{cao2023defending}
Cao, B., Cao, Y., Lin, L., and Chen, J.
\newblock Defending against alignment-breaking attacks via robustly aligned llm.
\newblock \emph{arXiv preprint arXiv:2309.14348}, 2023.

\bibitem[Chiang et~al.(2023)Chiang, Li, Lin, Sheng, Wu, Zhang, Zheng, Zhuang, Zhuang, Gonzalez, Stoica, and Xing]{vicuna2023}
Chiang, W.-L., Li, Z., Lin, Z., Sheng, Y., Wu, Z., Zhang, H., Zheng, L., Zhuang, S., Zhuang, Y., Gonzalez, J.~E., Stoica, I., and Xing, E.~P.
\newblock Vicuna: An open-source chatbot impressing gpt-4 with 90\%* chatgpt quality, March 2023.
\newblock URL \url{https://lmsys.org/blog/2023-03-30-vicuna/}.

\bibitem[Deng et~al.(2023)Deng, Zhang, Pan, and Bing]{deng2023multilingual}
Deng, Y., Zhang, W., Pan, S.~J., and Bing, L.
\newblock Multilingual jailbreak challenges in large language models.
\newblock \emph{arXiv preprint arXiv:2310.06474}, 2023.

\bibitem[Dettmers et~al.(2024)Dettmers, Pagnoni, Holtzman, and Zettlemoyer]{dettmers2024qlora}
Dettmers, T., Pagnoni, A., Holtzman, A., and Zettlemoyer, L.
\newblock Qlora: Efficient finetuning of quantized llms.
\newblock \emph{Advances in Neural Information Processing Systems}, 36, 2024.

\bibitem[Huang et~al.(2023)Huang, Gupta, Xia, Li, and Chen]{huang2023catastrophic}
Huang, Y., Gupta, S., Xia, M., Li, K., and Chen, D.
\newblock Catastrophic jailbreak of open-source llms via exploiting generation.
\newblock \emph{arXiv preprint arXiv:2310.06987}, 2023.

\bibitem[Hubinger et~al.(2024)Hubinger, Denison, Mu, Lambert, Tong, MacDiarmid, Lanham, Ziegler, Maxwell, Cheng, et~al.]{hubinger2024sleeper}
Hubinger, E., Denison, C., Mu, J., Lambert, M., Tong, M., MacDiarmid, M., Lanham, T., Ziegler, D.~M., Maxwell, T., Cheng, N., et~al.
\newblock Sleeper agents: Training deceptive llms that persist through safety training.
\newblock \emph{arXiv preprint arXiv:2401.05566}, 2024.

\bibitem[Inan et~al.(2023)Inan, Upasani, Chi, Rungta, Iyer, Mao, Tontchev, Hu, Fuller, Testuggine, et~al.]{inan2023llama}
Inan, H., Upasani, K., Chi, J., Rungta, R., Iyer, K., Mao, Y., Tontchev, M., Hu, Q., Fuller, B., Testuggine, D., et~al.
\newblock Llama guard: Llm-based input-output safeguard for human-ai conversations.
\newblock \emph{arXiv preprint arXiv:2312.06674}, 2023.

\bibitem[Jain et~al.(2023)Jain, Schwarzschild, Wen, Somepalli, Kirchenbauer, Chiang, Goldblum, Saha, Geiping, and Goldstein]{jain2023baseline}
Jain, N., Schwarzschild, A., Wen, Y., Somepalli, G., Kirchenbauer, J., Chiang, P.-y., Goldblum, M., Saha, A., Geiping, J., and Goldstein, T.
\newblock Baseline defenses for adversarial attacks against aligned language models.
\newblock \emph{arXiv preprint arXiv:2309.00614}, 2023.

\bibitem[Jiang et~al.(2023)Jiang, Sablayrolles, Mensch, Bamford, Chaplot, Casas, Bressand, Lengyel, Lample, Saulnier, et~al.]{jiang2023mistral}
Jiang, A.~Q., Sablayrolles, A., Mensch, A., Bamford, C., Chaplot, D.~S., Casas, D. d.~l., Bressand, F., Lengyel, G., Lample, G., Saulnier, L., et~al.
\newblock Mistral 7b.
\newblock \emph{arXiv preprint arXiv:2310.06825}, 2023.

\bibitem[Lees et~al.(2022)Lees, Tran, Tay, Sorensen, Gupta, Metzler, and Vasserman]{perspectiveapi}
Lees, A., Tran, V.~Q., Tay, Y., Sorensen, J.~S., Gupta, J., Metzler, D., and Vasserman, L.
\newblock A new generation of perspective api: Efficient multilingual character-level transformers.
\newblock \emph{Knowledge Discovery And Data Mining}, 2022.
\newblock \doi{10.1145/3534678.3539147}.

\bibitem[Lin et~al.(2023)Lin, Wang, Tong, Wang, Guo, Wang, and Shang]{lin2023toxicchat}
Lin, Z., Wang, Z., Tong, Y., Wang, Y., Guo, Y., Wang, Y., and Shang, J.
\newblock Toxicchat: Unveiling hidden challenges of toxicity detection in real-world user-{AI} conversation.
\newblock In \emph{The 2023 Conference on Empirical Methods in Natural Language Processing}, 2023.
\newblock URL \url{https://openreview.net/forum?id=jTiJPDv82w}.

\bibitem[Liu et~al.(2023)Liu, Xu, Chen, and Xiao]{liu2023autodan}
Liu, X., Xu, N., Chen, M., and Xiao, C.
\newblock Autodan: Generating stealthy jailbreak prompts on aligned large language models.
\newblock \emph{arXiv preprint arXiv:2310.04451}, 2023.

\bibitem[Mao et~al.(2023)Mao, Mohri, and Zhong]{mao2023cross}
Mao, A., Mohri, M., and Zhong, Y.
\newblock Cross-entropy loss functions: Theoretical analysis and applications.
\newblock \emph{arXiv preprint arXiv:2304.07288}, 2023.

\bibitem[Markov et~al.(2023)Markov, Zhang, Agarwal, Nekoul, Lee, Adler, Jiang, and Weng]{markov2023holistic}
Markov, T., Zhang, C., Agarwal, S., Nekoul, F.~E., Lee, T., Adler, S., Jiang, A., and Weng, L.
\newblock A holistic approach to undesired content detection in the real world.
\newblock In \emph{Proceedings of the AAAI Conference on Artificial Intelligence}, volume~37, pp.\  15009--15018, 2023.

\bibitem[Mehrotra et~al.(2023)Mehrotra, Zampetakis, Kassianik, Nelson, Anderson, Singer, and Karbasi]{mehrotra2023tree}
Mehrotra, A., Zampetakis, M., Kassianik, P., Nelson, B., Anderson, H., Singer, Y., and Karbasi, A.
\newblock Tree of attacks: Jailbreaking black-box llms automatically.
\newblock \emph{arXiv preprint arXiv:2312.02119}, 2023.

\bibitem[OpenAI(2023)]{openai2023gpt4}
OpenAI.
\newblock G{PT-4} technical report.
\newblock \emph{arXiv}, 2023.

\bibitem[Ouyang et~al.(2022)Ouyang, Wu, Jiang, Almeida, Wainwright, Mishkin, Zhang, Agarwal, Slama, Ray, et~al.]{ouyang2022training}
Ouyang, L., Wu, J., Jiang, X., Almeida, D., Wainwright, C., Mishkin, P., Zhang, C., Agarwal, S., Slama, K., Ray, A., et~al.
\newblock Training language models to follow instructions with human feedback.
\newblock \emph{Advances in Neural Information Processing Systems}, 35:\penalty0 27730--27744, 2022.

\bibitem[Qi et~al.(2023)Qi, Zeng, Xie, Chen, Jia, Mittal, and Henderson]{qi2023finetuning}
Qi, X., Zeng, Y., Xie, T., Chen, P.-Y., Jia, R., Mittal, P., and Henderson, P.
\newblock Fine-tuning aligned language models compromises safety, even when users do not intend to!, 2023.

\bibitem[Qi et~al.(2024)Qi, Zeng, Xie, Chen, Jia, Mittal, and Henderson]{anonymous2024finetuning}
Qi, X., Zeng, Y., Xie, T., Chen, P.-Y., Jia, R., Mittal, P., and Henderson, P.
\newblock Fine-tuning aligned language models compromises safety, even when users do not intend to!
\newblock In \emph{The Twelfth International Conference on Learning Representations}, 2024.
\newblock URL \url{https://openreview.net/forum?id=hTEGyKf0dZ}.

\bibitem[Qin et~al.(2022)Qin, Welleck, Khashabi, and Choi]{qin2022cold}
Qin, L., Welleck, S., Khashabi, D., and Choi, Y.
\newblock Cold decoding: Energy-based constrained text generation with langevin dynamics.
\newblock \emph{Advances in Neural Information Processing Systems}, 35:\penalty0 9538--9551, 2022.

\bibitem[Rebedea et~al.(2023)Rebedea, Dinu, Sreedhar, Parisien, and Cohen]{rebedea2023nemo}
Rebedea, T., Dinu, R., Sreedhar, M., Parisien, C., and Cohen, J.
\newblock Nemo guardrails: A toolkit for controllable and safe llm applications with programmable rails.
\newblock \emph{arXiv preprint arXiv:2310.10501}, 2023.

\bibitem[Robey et~al.(2023)Robey, Wong, Hassani, and Pappas]{robey2023smoothLLM}
Robey, A., Wong, E., Hassani, H., and Pappas, G.~J.
\newblock Smoothllm: Defending large language models against jailbreaking attacks.
\newblock \emph{arXiv preprint arXiv:2310.03684}, 2023.

\bibitem[Team et~al.(2023)Team, Anil, Borgeaud, Wu, Alayrac, Yu, Soricut, Schalkwyk, Dai, Hauth, et~al.]{team2023gemini}
Team, G., Anil, R., Borgeaud, S., Wu, Y., Alayrac, J.-B., Yu, J., Soricut, R., Schalkwyk, J., Dai, A.~M., Hauth, A., et~al.
\newblock Gemini: a family of highly capable multimodal models.
\newblock \emph{arXiv preprint arXiv:2312.11805}, 2023.

\bibitem[Touvron et~al.(2023{\natexlab{a}})Touvron, Martin, Stone, Albert, Almahairi, Babaei, Bashlykov, Batra, Bhargava, Bhosale, et~al.]{llama2}
Touvron, H., Martin, L., Stone, K., Albert, P., Almahairi, A., Babaei, Y., Bashlykov, N., Batra, S., Bhargava, P., Bhosale, S., et~al.
\newblock Llama 2: Open foundation and fine-tuned chat models.
\newblock \emph{arXiv preprint arXiv:2307.09288}, 2023{\natexlab{a}}.

\bibitem[Touvron et~al.(2023{\natexlab{b}})Touvron, Martin, Stone, Albert, Almahairi, Babaei, Bashlykov, Batra, Bhargava, Bhosale, et~al.]{touvron2023llama-2}
Touvron, H., Martin, L., Stone, K., Albert, P., Almahairi, A., Babaei, Y., Bashlykov, N., Batra, S., Bhargava, P., Bhosale, S., et~al.
\newblock Llama 2: Open foundation and fine-tuned chat models.
\newblock \emph{arXiv preprint arXiv:2307.09288}, 2023{\natexlab{b}}.

\bibitem[Wang et~al.(2023{\natexlab{a}})Wang, Chen, Pei, Xie, Kang, Zhang, Xu, Xiong, Dutta, Schaeffer, et~al.]{wang2023decodingtrust}
Wang, B., Chen, W., Pei, H., Xie, C., Kang, M., Zhang, C., Xu, C., Xiong, Z., Dutta, R., Schaeffer, R., et~al.
\newblock Decodingtrust: A comprehensive assessment of trustworthiness in gpt models.
\newblock \emph{NeurIPS}, 2023{\natexlab{a}}.

\bibitem[Wang et~al.(2023{\natexlab{b}})Wang, Liu, Park, Chen, and Xiao]{wang2023adversarial}
Wang, J., Liu, Z., Park, K.~H., Chen, M., and Xiao, C.
\newblock Adversarial demonstration attacks on large language models.
\newblock \emph{arXiv preprint arXiv:2305.14950}, 2023{\natexlab{b}}.

\bibitem[Wang et~al.(2018)Wang, Jha, and Chaudhuri]{wang2018analyzing}
Wang, Y., Jha, S., and Chaudhuri, K.
\newblock Analyzing the robustness of nearest neighbors to adversarial examples.
\newblock In \emph{International Conference on Machine Learning}, pp.\  5133--5142. PMLR, 2018.

\bibitem[Wei et~al.(2023)Wei, Wang, and Wang]{wei2023jailbreak}
Wei, Z., Wang, Y., and Wang, Y.
\newblock Jailbreak and guard aligned language models with only few in-context demonstrations.
\newblock \emph{arXiv preprint arXiv:2310.06387}, 2023.

\bibitem[Yang et~al.(2018)Yang, Qi, Zhang, Bengio, Cohen, Salakhutdinov, and Manning]{yang2018hotpotqa}
Yang, Z., Qi, P., Zhang, S., Bengio, Y., Cohen, W.~W., Salakhutdinov, R., and Manning, C.~D.
\newblock Hotpotqa: A dataset for diverse, explainable multi-hop question answering.
\newblock \emph{arXiv preprint arXiv:1809.09600}, 2018.

\bibitem[Yong et~al.(2023)Yong, Menghini, and Bach]{yong2023low}
Yong, Z.-X., Menghini, C., and Bach, S.~H.
\newblock Low-resource languages jailbreak gpt-4.
\newblock \emph{arXiv preprint arXiv:2310.02446}, 2023.

\bibitem[Yuan et~al.(2023)Yuan, Jiao, Wang, Huang, He, Shi, and Tu]{yuan2023gpt}
Yuan, Y., Jiao, W., Wang, W., Huang, J.-t., He, P., Shi, S., and Tu, Z.
\newblock Gpt-4 is too smart to be safe: Stealthy chat with llms via cipher.
\newblock \emph{arXiv preprint arXiv:2308.06463}, 2023.

\bibitem[Zeng et~al.(2024)Zeng, Lin, Zhang, Yang, Jia, and Shi]{zeng2024johnny}
Zeng, Y., Lin, H., Zhang, J., Yang, D., Jia, R., and Shi, W.
\newblock How johnny can persuade llms to jailbreak them: Rethinking persuasion to challenge ai safety by humanizing llms.
\newblock \emph{arXiv preprint arXiv:2401.06373}, 2024.

\bibitem[Zhang et~al.(2023)Zhang, Shen, Tao, Cheng, and Zhang]{zhang2023make}
Zhang, Z., Shen, G., Tao, G., Cheng, S., and Zhang, X.
\newblock Make them spill the beans! coercive knowledge extraction from (production) llms.
\newblock \emph{arXiv preprint arXiv:2312.04782}, 2023.

\bibitem[Zheng et~al.(2023)Zheng, Chiang, Sheng, Zhuang, Wu, Zhuang, Lin, Li, Li, Xing, et~al.]{zheng2023judging}
Zheng, L., Chiang, W.-L., Sheng, Y., Zhuang, S., Wu, Z., Zhuang, Y., Lin, Z., Li, Z., Li, D., Xing, E., et~al.
\newblock Judging llm-as-a-judge with mt-bench and chatbot arena.
\newblock \emph{arXiv preprint arXiv:2306.05685}, 2023.

\bibitem[Zou et~al.(2023)Zou, Wang, Kolter, and Fredrikson]{zou2023universal}
Zou, A., Wang, Z., Kolter, J.~Z., and Fredrikson, M.
\newblock Universal and transferable adversarial attacks on aligned language models.
\newblock \emph{arXiv preprint arXiv:2307.15043}, 2023.

\end{thebibliography}
\bibliographystyle{icml2024}

\newpage
\appendix
\onecolumn

\section*{Appendix}

\section{Per-Category Performance}
\label{app:category-wise}
\begin{figure}[!h]
    \centering
    \includegraphics[width=0.7\linewidth]{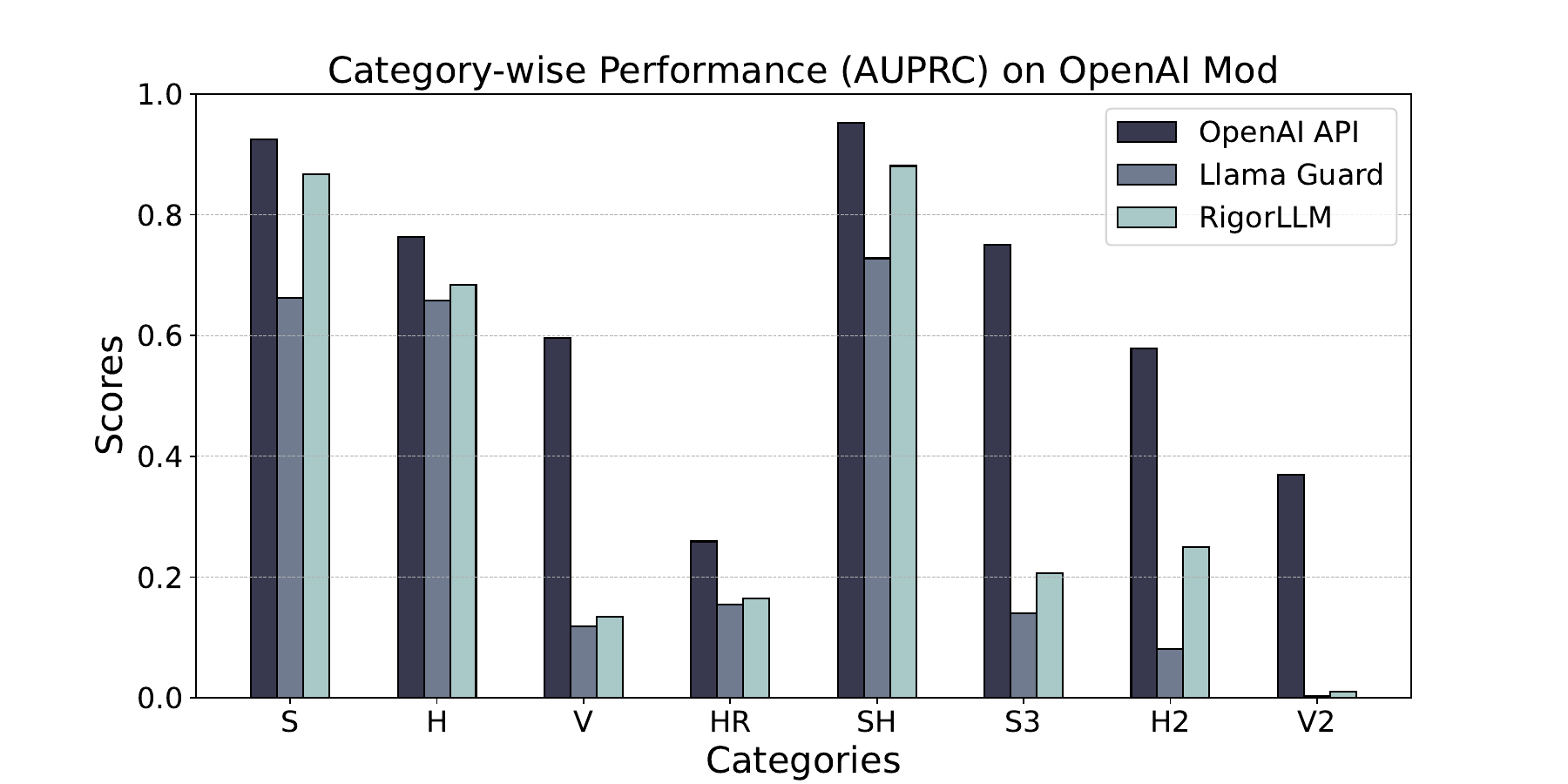}
    \caption{Category-Wise Performance on OpenAI Moderation Dataset.}
    \label{fig:category-wise}
\end{figure}

We report the per-category results for OpenAI API, LlamaGuard and \name on the OpenAI Moderation Dataset in Figure~\ref{fig:category-wise}.
For \name, we map the categories of its training data to those of the OpenAI Moderation Dataset to calculate the category-based content moderation results for comparison.
We observe that although \name achieves much higher performance than the OpenAI API overall, it is challenging to compare by category since there is a mismatch between the taxonomies of risks. However, \name can still significantly outperform LlamaGuard on each category.

\section{Additional Ablation Studies}

\subsection{Scaling Law}
\label{app:scale}
\begin{figure}[h!]
    \centering
    \subfigure[AUPRC vs Number of Generated Data per Category]{%
        \includegraphics[width=0.4\textwidth]{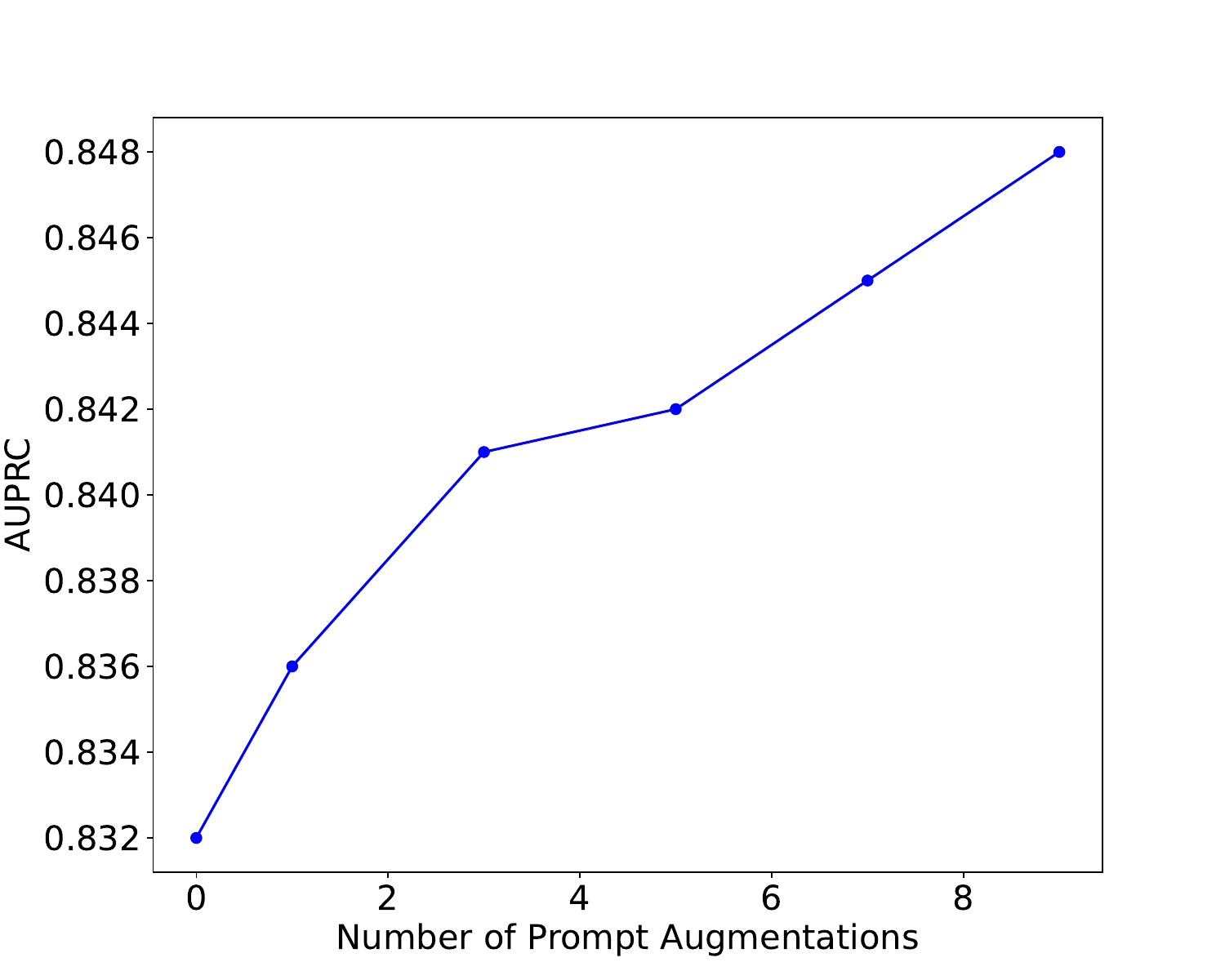}%
        \label{fig:generated_data}%
    }
    \subfigure[AUPRC vs Number of Prompt Augmentations]{%
        \includegraphics[width=0.4\textwidth]{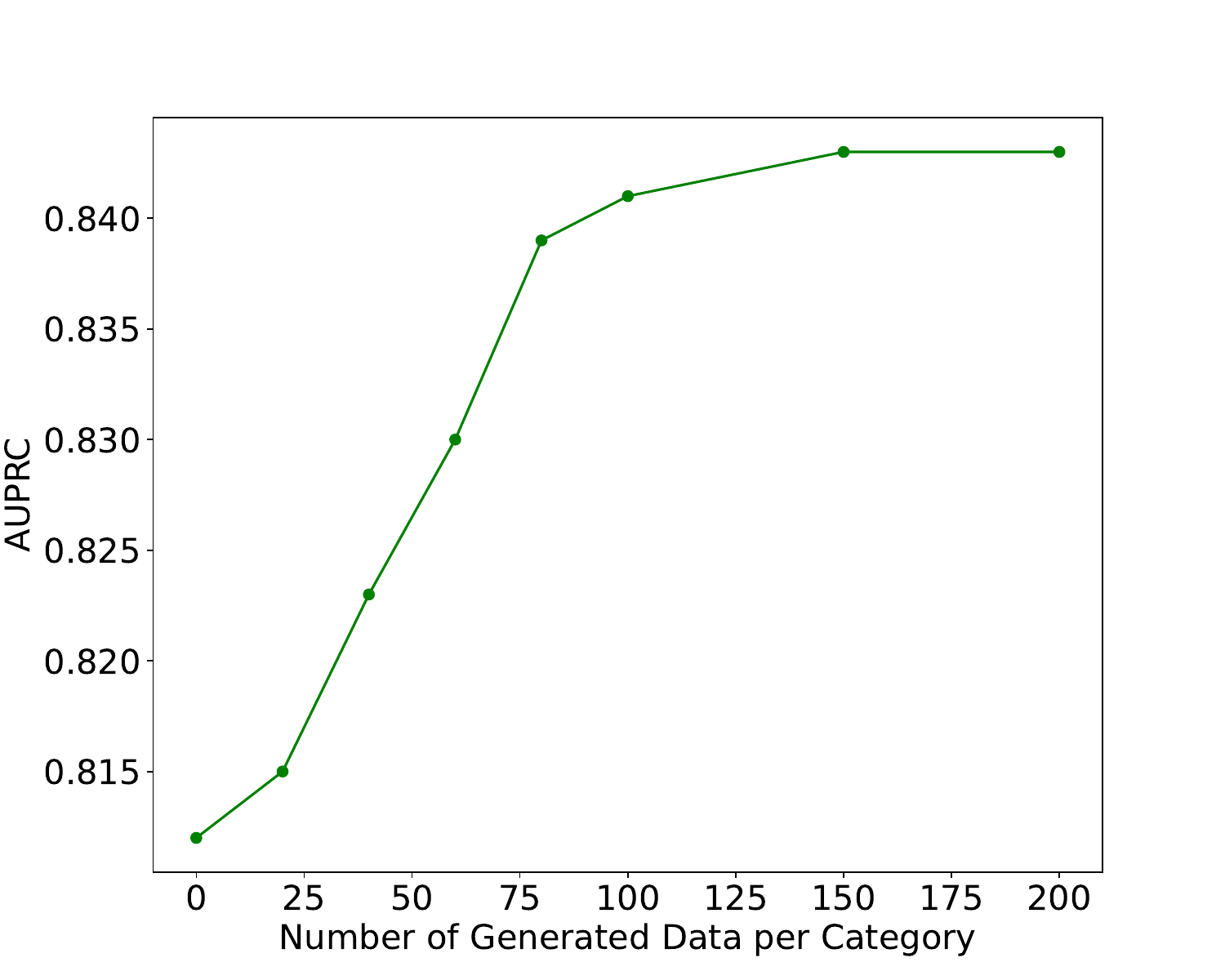}%
        \label{fig:augmentations}%
    }
    
    \caption{Comparison of AUPRC under different numbers of generated data during energy-based data generation and the number of prompt augmentations.}
    \label{fig:comparison}
\end{figure}

In Figure~\ref{fig:generated_data} and Figure~\ref{fig:augmentations}, we report the scaling law of \name on different numbers of generated data during energy-based data generation and the number of generated paraphrases and summaries during prompt augmentation. We can see that with an increased number of generated training data and prompt augmentations, the performance of \name can be further increased.

\subsection{Probability Distribution Across Different Datasets}
\begin{figure}[!h]
    \centering
    \includegraphics[width=0.42\linewidth]{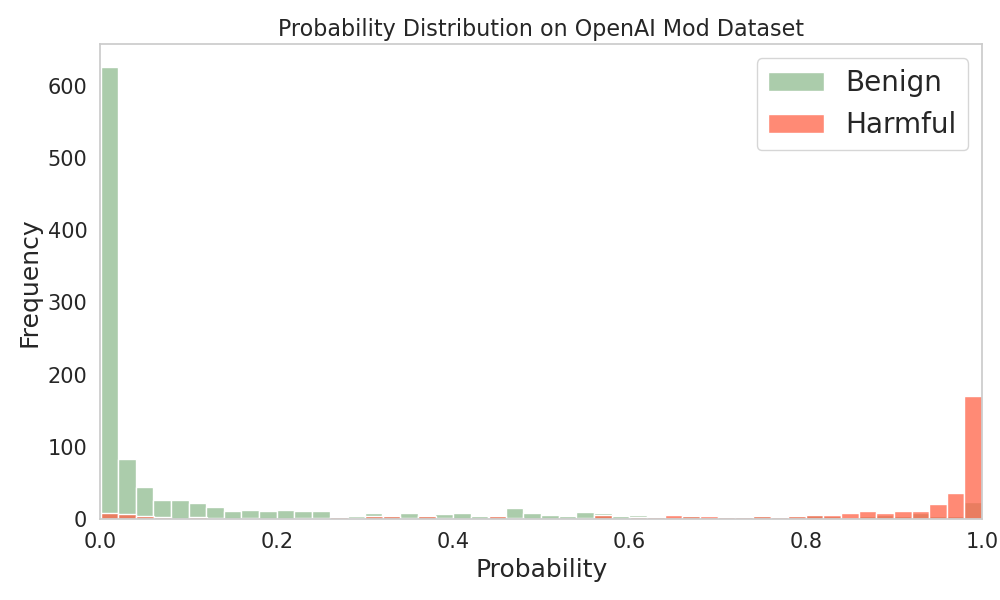}
    \includegraphics[width=0.42\linewidth]{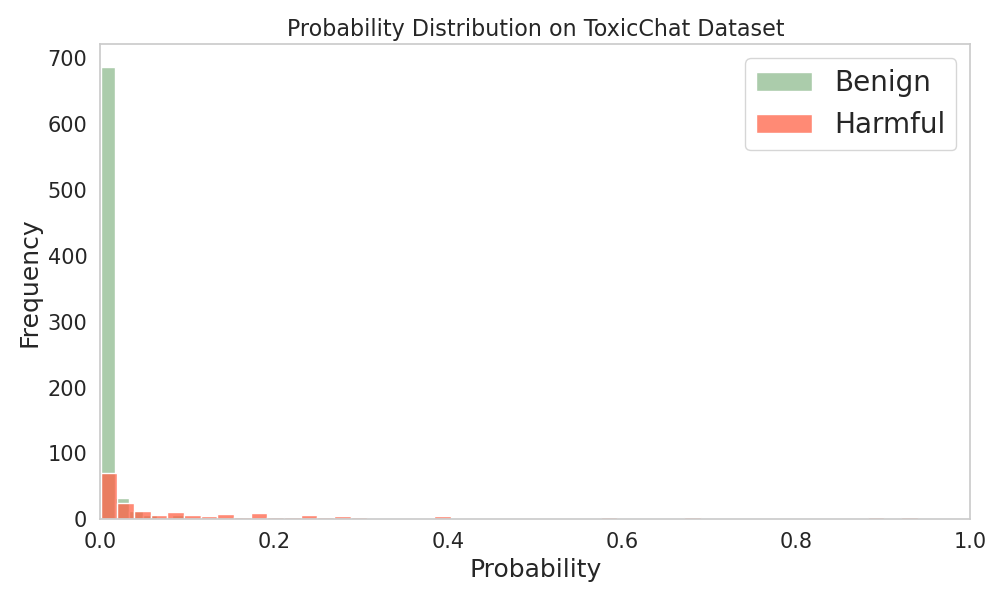}
    \includegraphics[width=0.42\linewidth]{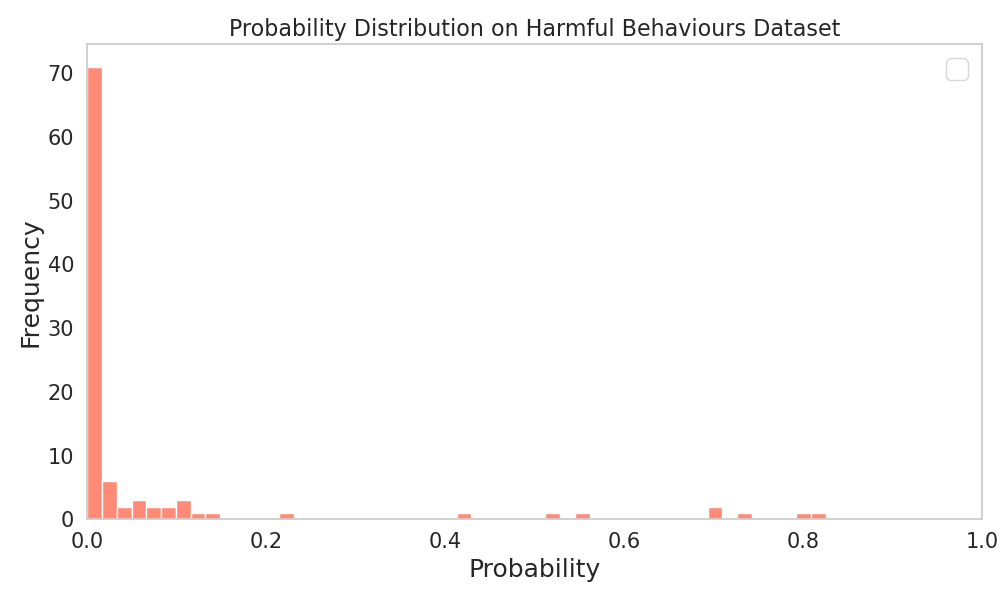}
    \includegraphics[width=0.42\linewidth]{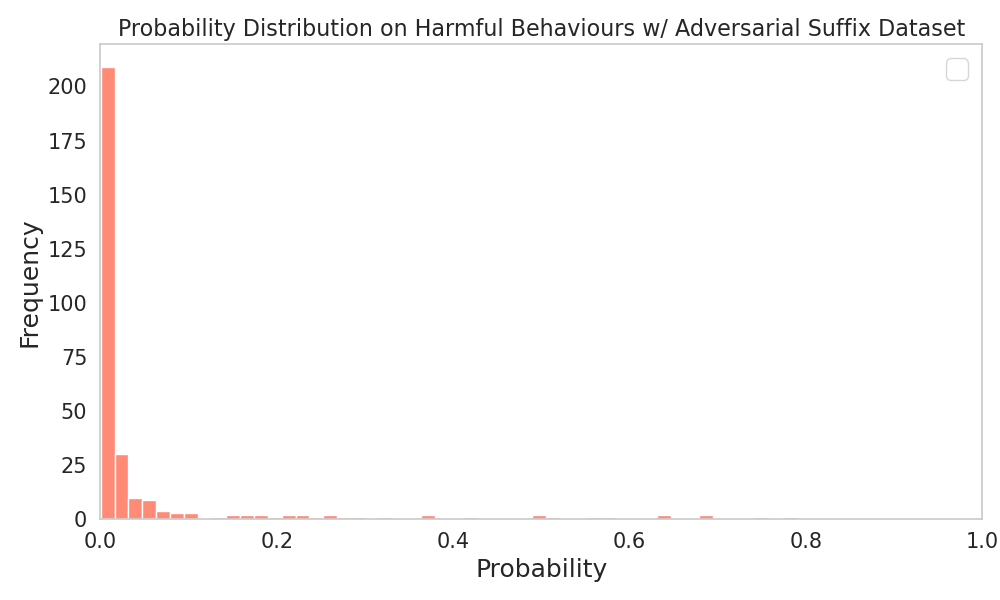}
    \caption{Probability distribution of OpenAI API across different datasets.}
    \label{fig:openai_distribution}
\end{figure}

\begin{figure}[!h]
    \centering   
    \includegraphics[width=0.42\linewidth]{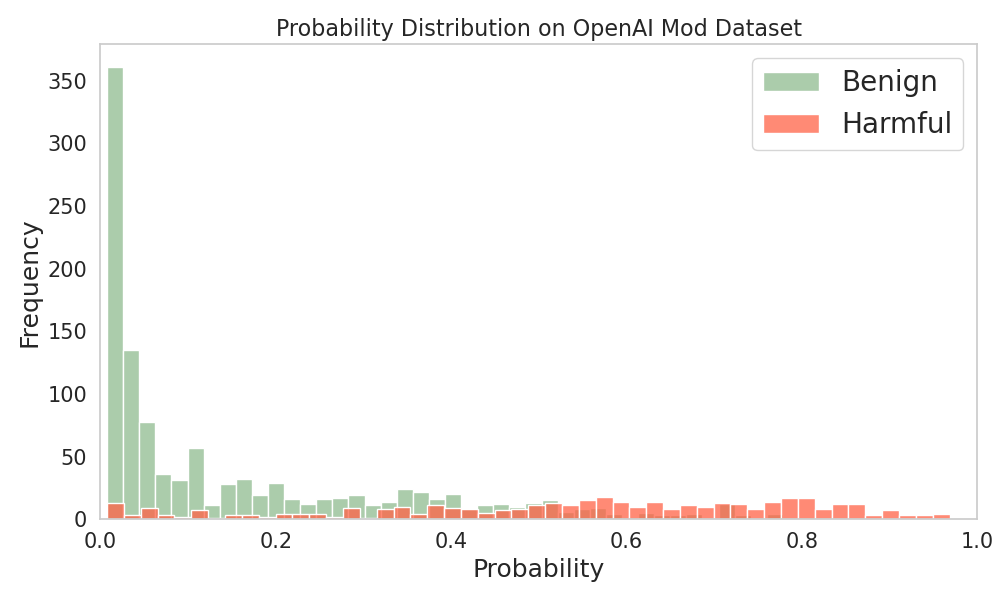}
    \includegraphics[width=0.42\linewidth]{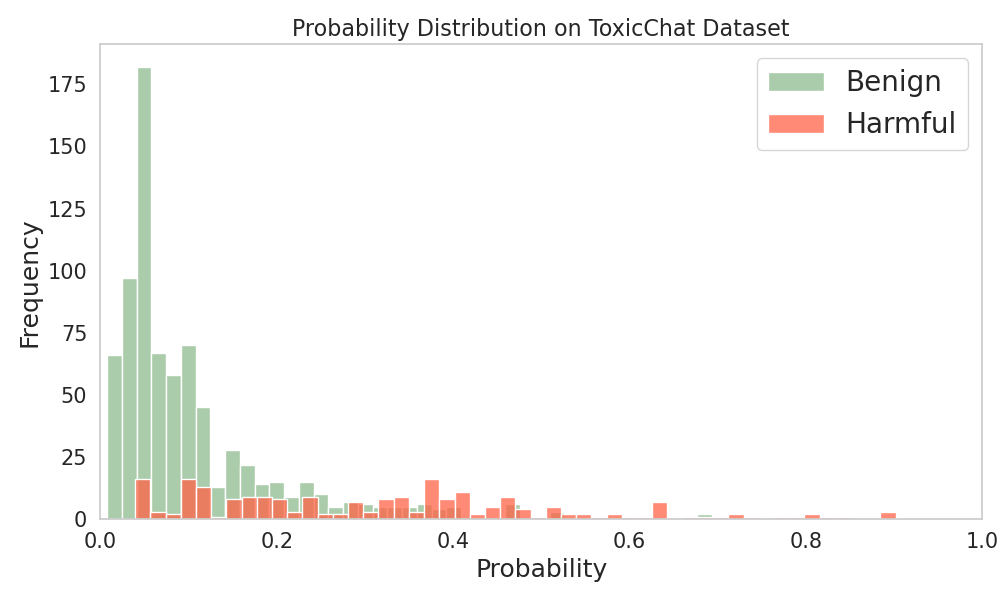}
    \includegraphics[width=0.42\linewidth]{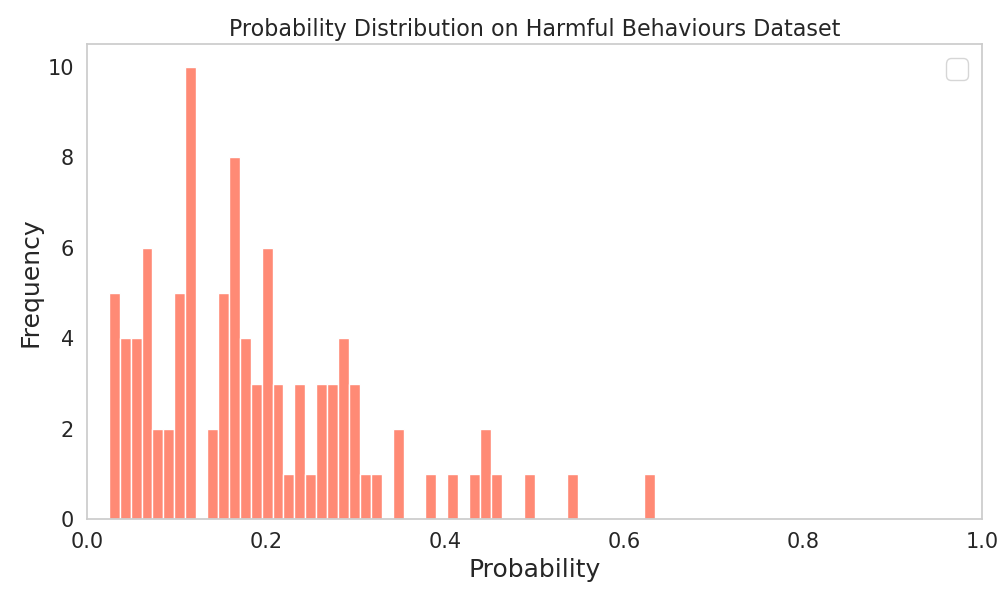}
    \includegraphics[width=0.42\linewidth]{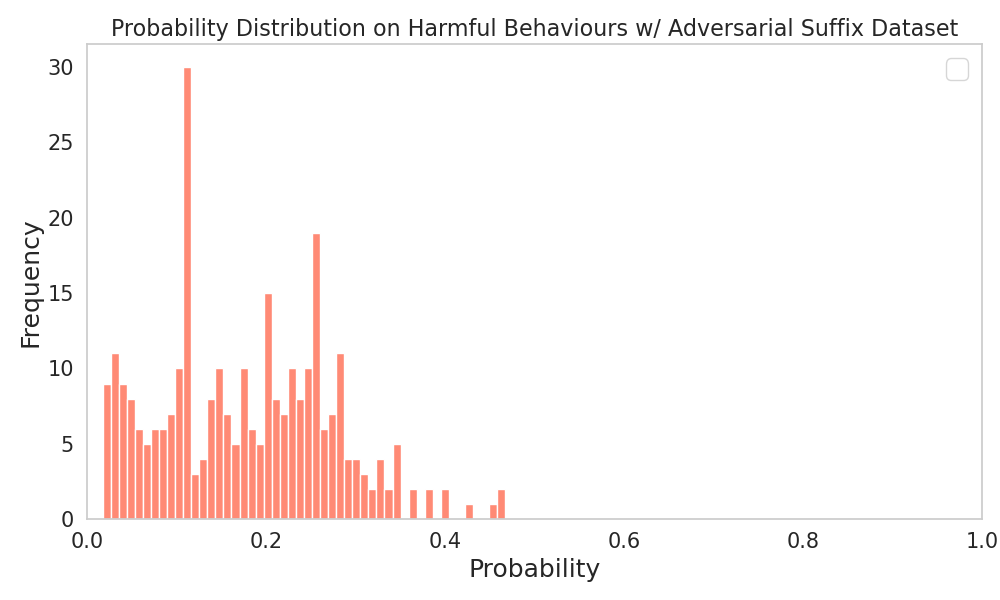}
    \caption{Probability distribution of Perspective API across different datasets.}
    \label{fig:perspective_distribution}
\end{figure}

In Figure~\ref{fig:openai_distribution} and Figure~\ref{fig:perspective_distribution}, we plot out the distributions of probabilities for OpenAI API and Perspective API, respectively. We can see that for both baselines, the predictions are concentrated on the low-probability region, indicating that they fail to detect the harmful inputs.

\label{app:distribution}

\end{document}